\documentstyle[sprocl,epsf]{article}
\input{psfig}
\def\Journal#1#2#3#4{{#1} {\bf #2}, #3 (#4)}

\def\NPB{{\em Nucl. Phys.} B}

\def\PRL{{\em Phys. Rev. Lett.}}
\def\PRD{{\em Phys. Rev.} D}
\def\ZPC{{\em Z. Phys.} C}

\def\be{\begin{equation}}
\def\ee{\end{equation}}
\def\bea{\begin{eqnarray}}
\def\eea{\end{eqnarray}}

\begin{document}
\vspace*{-1.5 in}
{\flushright {FERMILAB-CONF-96/235-E}}
\vspace*{1.0in}
\title{CHARMONIUM AND BOTTOMONIUM PRODUCTION IN $p\overline{p}$
COLLISIONS AT CDF}
\author{ MARK W. BAILEY \footnote{Representing the CDF Collaboration}}
\address{The University of New Mexico, Albuquerque, NM 87131}
\maketitle\abstracts{
We present measurements of charmonium and bottomonium production using a data
sample collected by CDF during the 1992-93 $p\overline{p}$ collider run at the
Fermilab Tevatron.}
\section{Introduction}
  We present measurements of charmonium and bottomonium production using a data
sample collected by CDF during the 1992-93 $p\overline{p}$ collider run at the
Fermilab Tevatron.

   Previous CDF measurements of $J/\psi$ and $\psi(2S)$ production during the
1988-89 collider run showed production cross sections considerably larger than
contemporary theory predicted.  This drew theoretical interest, but at the time
the question of  whether or not the excess could be attributed to a large
prompt component was not addressed.

   In these analyses, differential cross-sections for $J/\psi$, $\psi(2S)$ and
three $\Upsilon$ states have been measured using the $\mu^+\mu^-$ decay
channel.  Using a silicon vertex detector, we separate the $J/\psi$ and
$\psi(2S)$ samples into their components arising from $b$ decay and from prompt
production by analyzing the proper decay length distributions.
Additionally,  by reconstruction of the decay 
$\chi_c \rightarrow J/\psi \gamma$, we
determine the fraction of the prompt $J/\psi$ sample coming from $\chi_c$
decay, compared to that from direct charm and gluon fragmentation.

   We also briefly summarize recent proposed improvements in the theoretical
description of the observed production {rates.\cite{braatrev}}

\section{Charmonium total cross sections}

      $\psi(2S)$ and $J/\psi$ candidates are reconstructed via their dimuon
decay modes.  Each muon is required to be detected in the CDF central muon
chambers, which cover the pseudorapidity range $|\eta|<0.6$,  and to have
transverse momentum, $p_T$, greater than 2.0 GeV$/c$.  Additionally, at least
one muon is required to have $p_T > 2.8$ GeV/$c$, and the $J/\psi$ or
$\psi(2S)$ candidate must have $p_T > 5.0$ GeV$/c$.  About 22,000 $J/\psi$ and
800 $\psi(2S)$ are reconstructed in data samples of 15.4
pb$^{-1}$ and 17.8 pb$^{-1}$, respectively.  The product of dimuon branching
ratio times integrated cross section is found to be 
\mbox {$17.35\pm0.14\pm2.79$ nb} for
$J/\psi$ and \mbox{$0.57\pm0.04\pm0.09$ nb} for $\psi(2S)$.

\section{Charmonium from $b$ Decay}

\begin{figure} 
\leftline{
\epsfysize 5.5cm
\epsffile{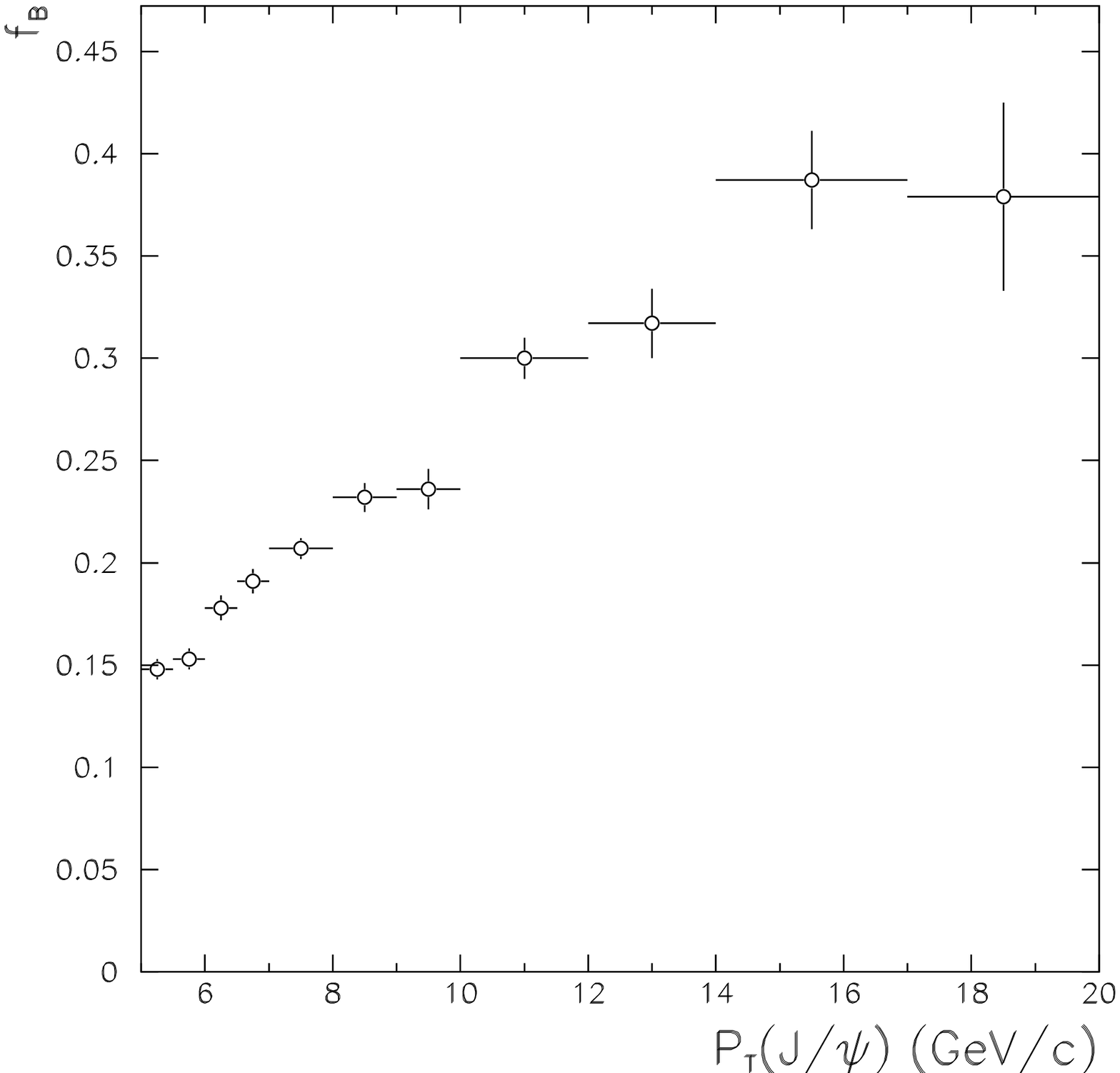}
\epsfysize 5.5cm 
\epsffile{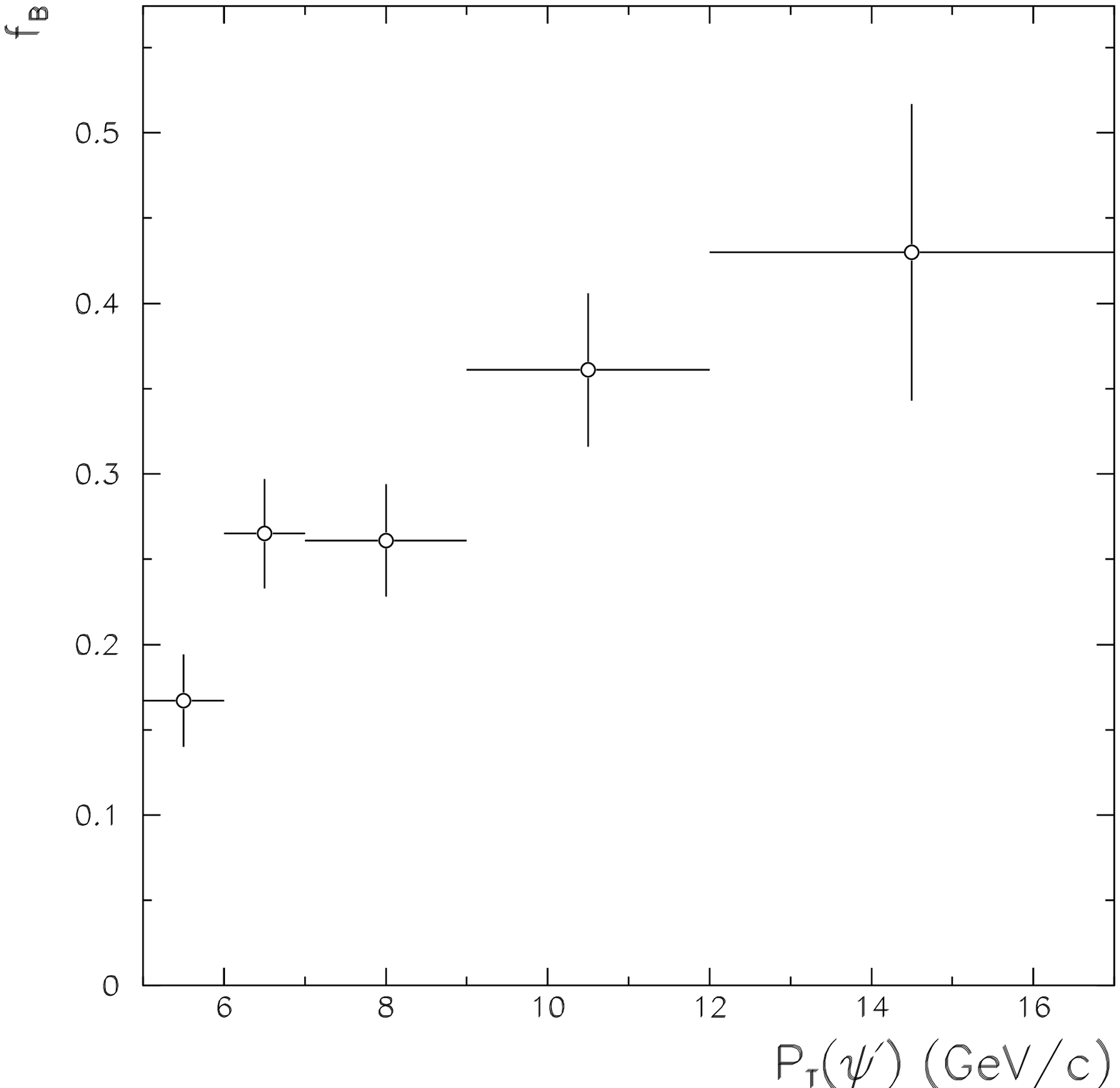} }
\caption{Fraction of $J/\psi$ (left) and $\psi(2S)$ (right) from $b$ decay 
as a function of $p_T$.
\label{myfig1}} 
\end{figure}

\begin{figure} 
\leftline{
\epsfysize 5.5cm
\epsffile{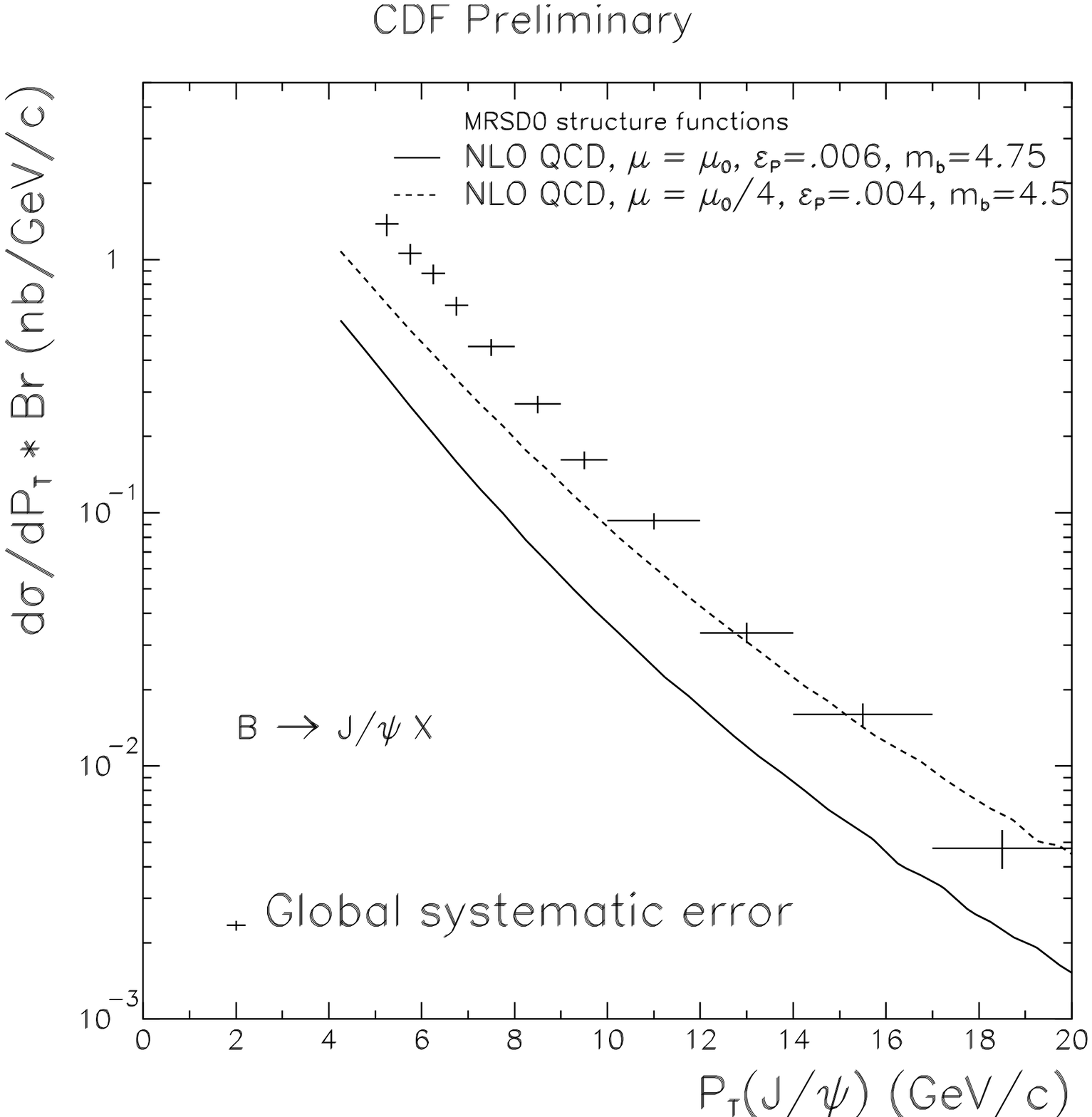}
\epsfysize 5.5cm 
\epsffile{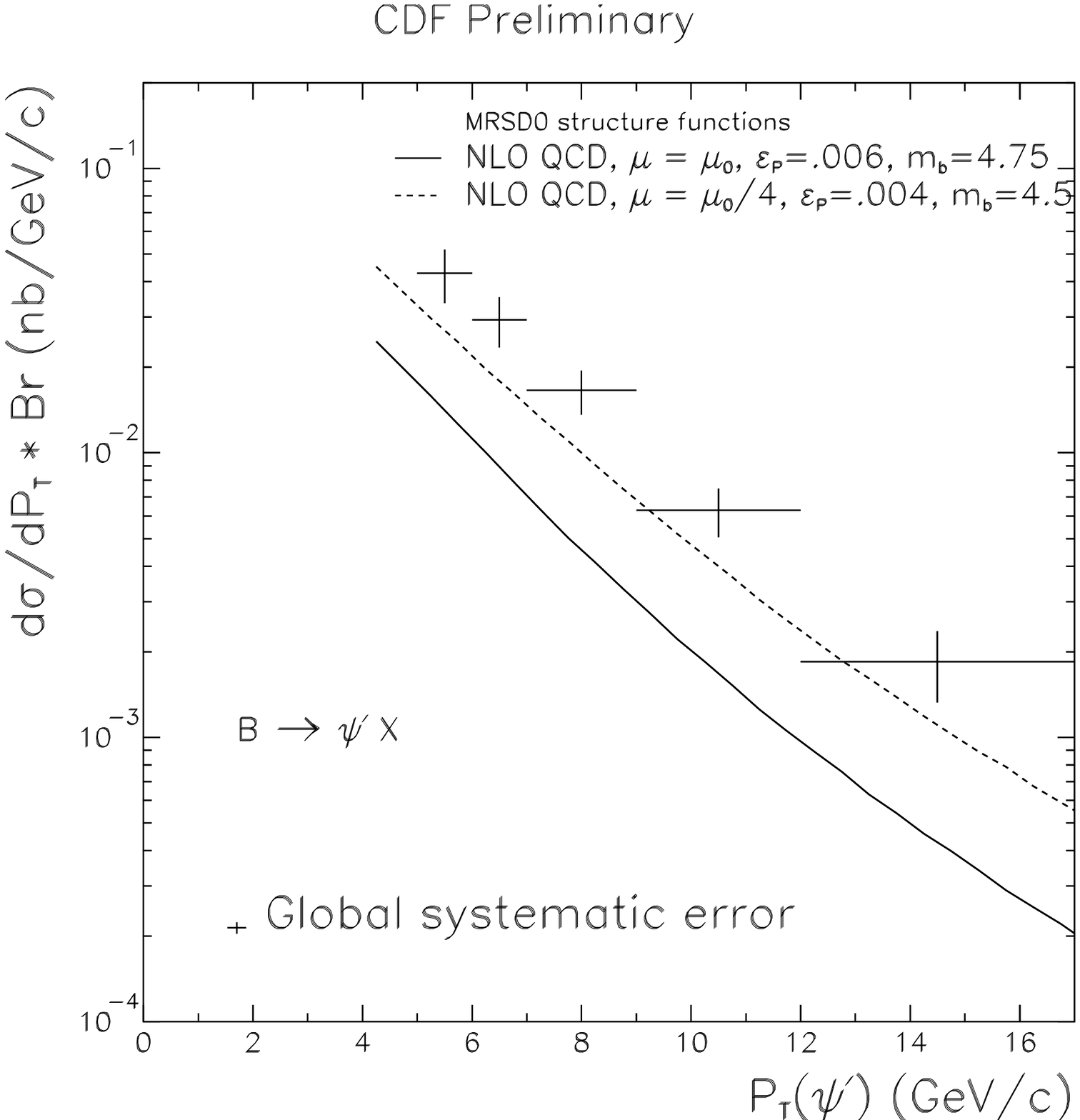} }
\caption{$b$ quark cross section determined from the $J/\psi$ (left) and
$\psi(2S)$ (right) samples.
\label{myfig3}}
\end{figure}

     The fraction of $\psi(2S)$ and $J/\psi$ resulting from the decay of $b$
hadrons is determined by an analysis of the proper decay length
($c\tau$) distributions.  The $c\tau$ distribution is fitted to three
components:  an exponential 
convoluted with a Gaussian resolution function for the $b$ hadron decay
component, a Gaussian function centered at 0 for prompt production, and a 
Gaussian function with
positive and negative exponential tails to described the background, both
combinatorial 
as well as from, e.g., sequential $b\rightarrow\mu^- c \rightarrow \mu^+ s$
decays.  The samples are subdivided into ranges of $p_T(\mu^+\mu^-)$ and
fitted separately for each range.  Figure \ref{myfig1}
 shows the fraction from $b$ decay as a function of $p_T$ for 
$J/\psi$ and $\psi(2S)$.  These fractions are then convoluted
with the charmonia $p_T$ spectra to give the $b$ cross section, as shown in
figure \ref{myfig3}.  The results are within a factor of 2-3
of the NLO QCD \mbox{prediction,\cite{nde}} using a central value of the 
input parameters, and are
consistent with other CDF $b$ cross section \mbox{results.\cite{myprl}}

\section{Prompt Charmonium Production}

    The large fraction of prompt charmonium production is in disagreement
with theoretical predictions based on color singlet production of
$c\overline{c}$ bound \mbox{states.\cite{gms}}  The rate of prompt $\psi(2S)$
production is about a factor of 50 larger than predictions based on such
a model.  All prompt $\psi(2S)$ are believed to be directly produced, since
$\chi_c$ states with sufficient mass to decay to $\psi(2S)$ lie above the
threshold for strong decays to $D\overline{D}$ meson pairs.  However, prompt
$J/\psi$ are produced not only directly, but also via radiative decays $\chi_c
\rightarrow J/\psi \gamma$.  

   To determine the fraction of prompt {\em
direct} $J/\psi$ produced, we fully reconstruct $\chi_c$ states.
Photon candidates detected in the central electromagnetic calorimeter with
energy greater than 1 GeV and having no charged track pointing to the same
calorimeter tower are combined with $\mu^+\mu^-$, and a peak containing
$1230\pm72$ $\chi_c$ candidates is
observed in the mass difference
distribution $M(\mu^+\mu^-\gamma)-M(\mu^+\mu^-)$. 
The background under the peak has been modelled
by embedding simulated $\pi^0$ and $\eta^0$ decay photons in real $J/\psi$
events.  The fraction of $J/\psi$ coming from prompt $\chi_c$ decay,
measured for 4
different $p_T$ bins, ranges from about $32\%$ for 4-6 GeV/$c$ to
$28\%$ for $>10$ GeV/$c$.  Multiplying the total prompt  $J/\psi$ 
cross section by the $\chi_c$ fraction shows that the rate of $J/\psi$
production from $\chi_c$ is within a factor of 2-3 of the theoretical
prediction, but, as with the $\psi(2S)$, the remaining direct $J/\psi$ cross
section is about a factor of 50 larger than the color singlet prediction.

\begin{figure} 
\leftline{
\epsfysize 5.5cm
\epsffile{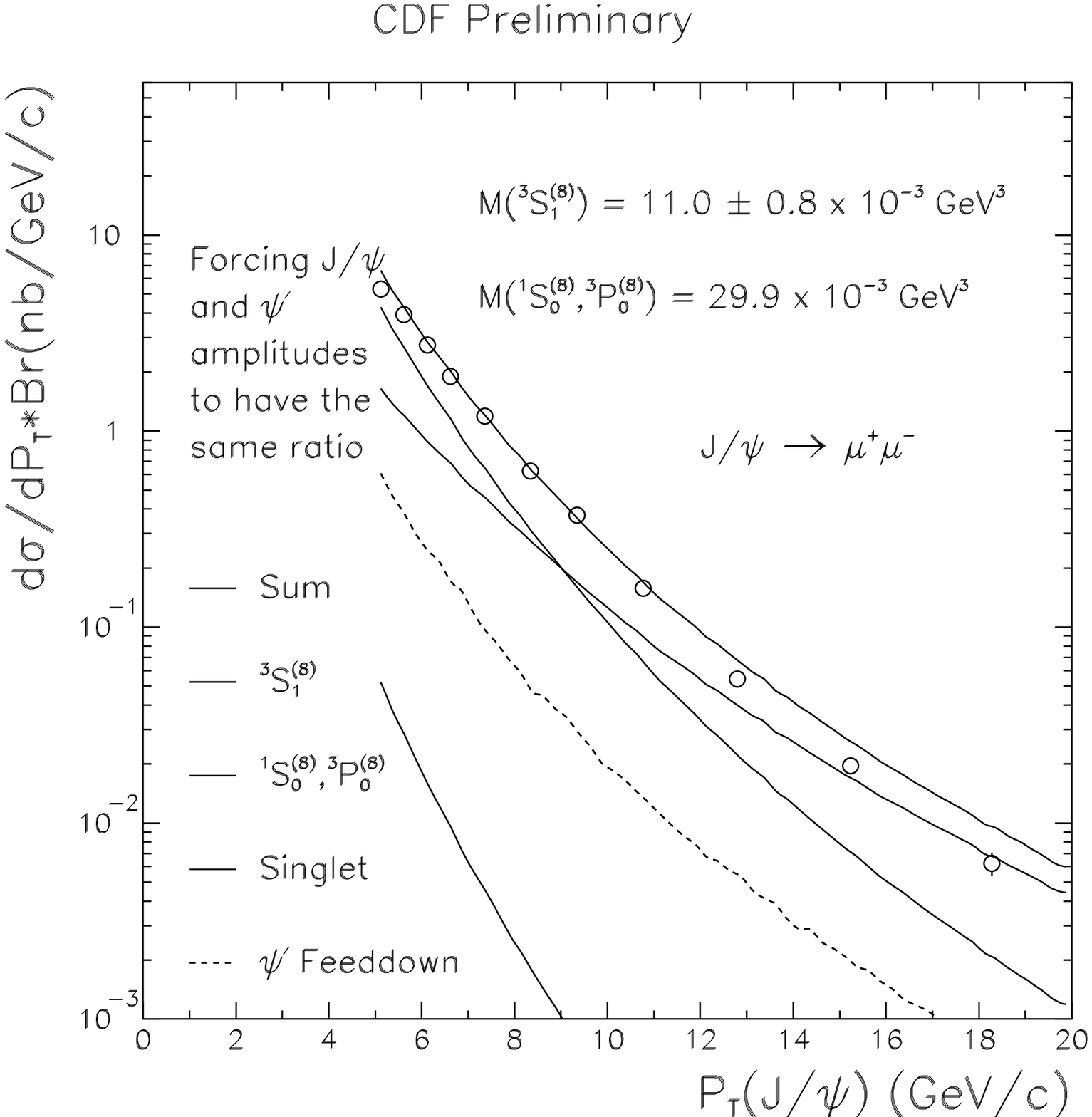}
\epsfysize 5.5cm 
\epsffile{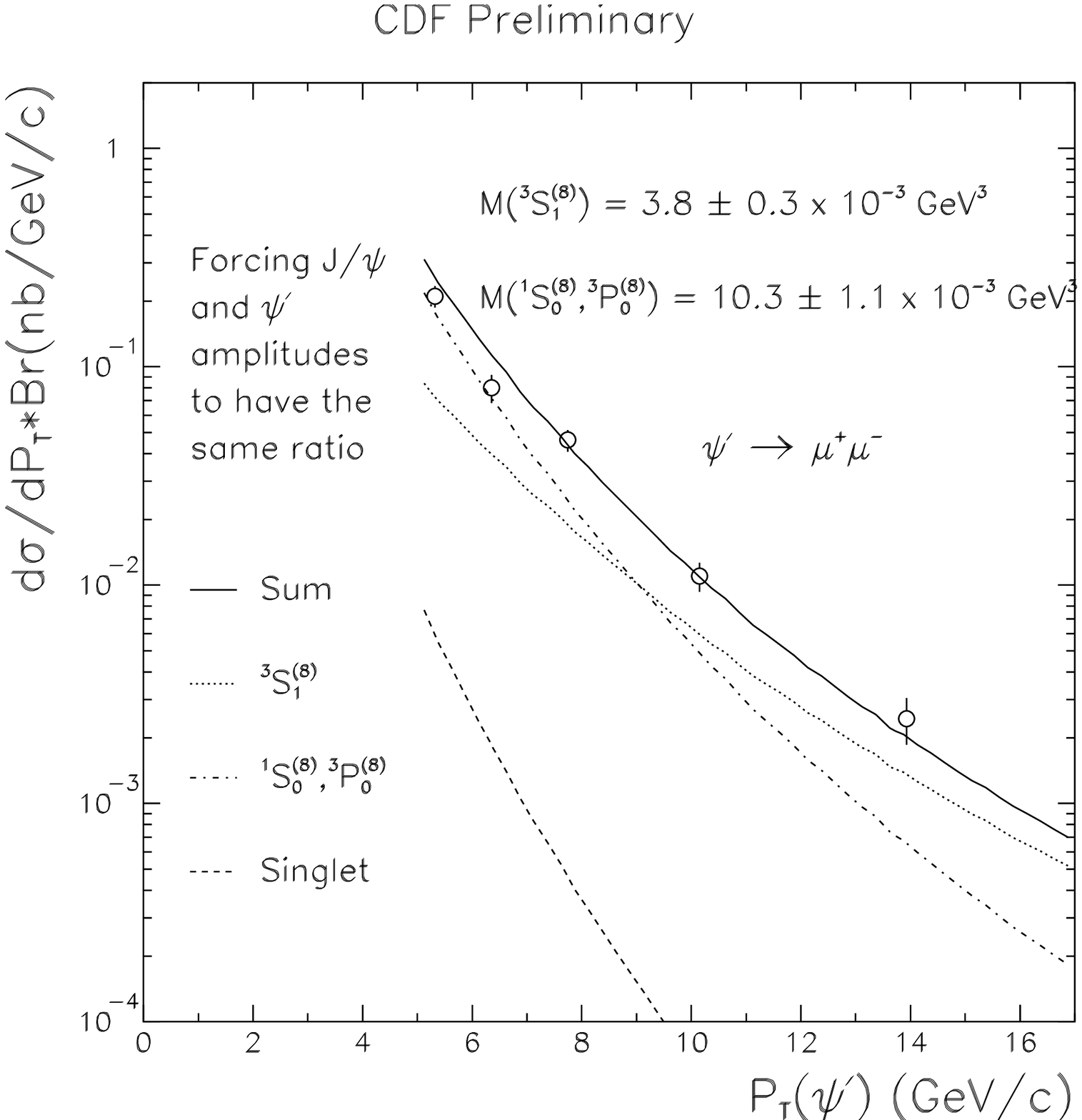} }
\caption{Prompt direct $J/\psi$ (left) and $\psi(2S)$ (right)
cross-sections, compared to the theoretical
prediction, with color octet components fitted simultaneously to the
$J/\psi$ and $\psi(2S)$ distributions.
\label{myfig5}}
\end{figure}

     One proposal to explain the prompt charmonium production rates observed is
to include $c\overline{c}$ pairs produced in a color octet
\mbox{state.\cite{braflem}}  The initial
production can be calculated perturbatively and can be used to predict the
$p_T$ dependence of the cross section.  The transition to a color singlet
state needed to form a bound $c\overline{c}$ particle proceeds via soft gluon
emission.  This latter process cannot be calculated perturbatively, so the
normalization is found by fitting the theory to the data.  Figure \ref{myfig5}
shows the prompt $J/\psi$ and $\psi(2S)$ cross sections and the
corresponding theoretical predictions when the fitted color octet contributions
are \mbox{included.\cite{cho}}  Further cross-checks in hadro- and
photo-production experiments are desirable, as is a measurement CDF hopes to
make of the $\psi(2S)$ polarization, which is predicted to be transversely
polarized in this model.

\section{Bottomium Production}

     CDF has also published production cross sections for 
$\Upsilon(1S)$, $\Upsilon(2S)$, and $\Upsilon(3S)$ based on a data
sample of \mbox{16 pb$^{-1}$.\cite{upsohl}}   All three states combined yield a
total of about 1800 
candidates reconstructed in the $\mu^+\mu^-$ decay channel.

\begin{figure} 
\epsfysize 2.5in
\epsffile{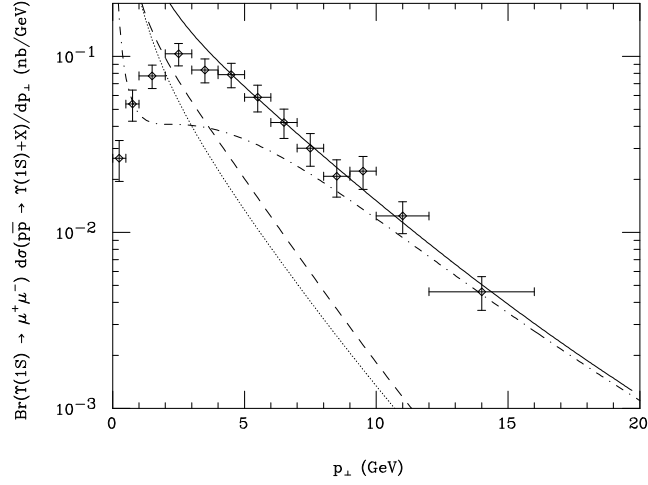}
\caption{$\Upsilon(1S)$
differential cross section, compared to the theoretical
prediction.  The dotted line shows the color singlet contribution,
while the dashed lines are the color octet components fitted to the data.
The solid line is the sum of all contributions.
\label{myfig7}}
\end{figure}

\begin{figure} 
\epsfysize 2.5in
\epsffile{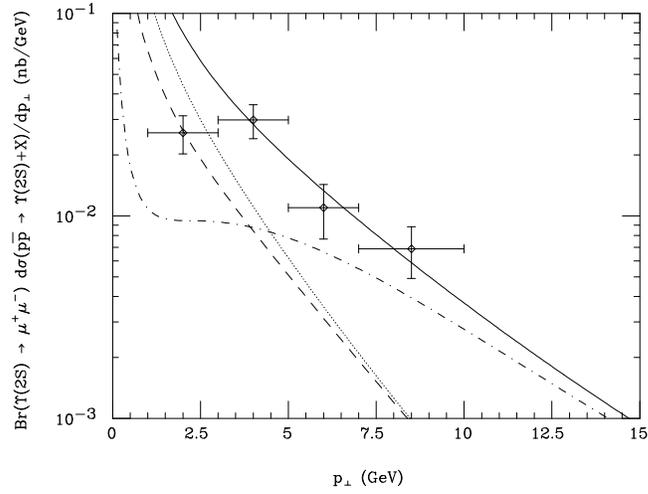}
\caption{$\Upsilon(2S)$
differential cross section, compared to the theoretical
prediction.  The dotted line shows the color singlet contribution,
while the dashed lines are the color octet components fitted to the data.
The solid line is the sum of all contributions.
\label{myfig8}}
\end{figure}

     Recently, a theoretical prediction including color octet contributions was
fitted to the $\Upsilon(1S)$ and $\Upsilon(2S)$ differential 
\mbox{distributions.\cite{cho}}  Figures \ref{myfig7} and \ref{myfig8} show the
fit results, which provide good descriptions of the shapes of the
distributions.

     Increased statistics using CDF's full 110-pb$^{-1}$ data sample will allow
finer binning in $p_T$, improving the experimental description of the shape. 
Reconstruction of the $\chi_b$ states via $\Upsilon\gamma$ decay --- while
difficult due to the small mass difference between $\chi_b$ and $\Upsilon$ ---
would provide an additional probe of the underlying production mechanisms.

\section{Conclusion}

     CDF has measured the differential production cross sections for $J/\psi$,
$\psi(2S)$, and three $\Upsilon$ states.  The prompt and $b$-decay parts of
both charmonium states have been extracted.  The $J/\psi$ prompt cross section
has been further subdivided into its direct and $\chi_c$ components.

     These measurements provided the impetus for improved theoretical models,
such as the color octet model, which show potential to explain charmonium
production in $p\overline{p}$ collisions.  Additional experimental results,
such as measurement of the $\psi(2S)$ polarization and reconstruction of
$\chi_b$ states --- both possible with CDF's existing dataset --- may provide
additional insight into the underlying production mechanisms.


\section*{Acknowledgments}      We thank the Fermilab staff and the technical
staffs of the  
participating institutions for their vital contributions.  This work was
supported by the U.S. Department of Energy and National Science Foundation;
the Italian Istituto Nazionale di Fisica Nucleare; the Ministry of Education,
Science and Culture of Japan; the Natural Sciences and Engineering Research
Council of Canada; the National Science Council of the Republic of China;
the A. P. Sloan Foundation; and the Alexander von Humboldt-Stiftung.

\end{document}